\documentclass[preprint2]{aastex}
\usepackage{psfig}
\newcommand{\lta}{$\; \buildrel < \over \sim \;$}
\newcommand{\simlt}{\lower.5ex\hbox{\lta}}
\newcommand{\gta}{$\; \buildrel > \over \sim \;$}
\newcommand{\simgt}{\lower.5ex\hbox{\gta}}

\newcommand{\kms}{{\rm\,km\,s^{-1}}}
\newcommand{\pc}{{\rm\,pc}}

\newcommand{\msun}{{\rm\,M_\odot}}
\newcommand{\lsun}{{\rm\,L_\odot}}

\newcommand{\ffffff}[1]{\mbox{$#1$}}

\newcommand{\scmd}{\mbox{\ffffff{''}}}

\slugcomment{Accepted for Publication in the Astrophysical Journal}

\begin{document}

\title{The Search for Cosmological Black Holes: \\
A Surface Brightness Variability Test}

\author{
Geraint F. Lewis\altaffilmark{1,2} \& Rodrigo A. Ibata\altaffilmark{3}
\altaffiltext{1}{ 
Astronomy Dept., University of Washington, \\ 
Seattle WA, U.S.A. \& \\ 
Dept. of Physics and Astronomy, University of Victoria, 
Victoria, B.C., Canada \\
Electronic mail: {\tt gfl@astro.washington.edu} \\
Electronic mail: {\tt gfl@uvastro.phys.uvic.ca}
}
\altaffiltext{2}
{Present Address: Anglo-Australian Observatory, \\ PO Box 296, 
Epping, NSW 1710, Australia \\
Electronic mail: {\tt gfl@aaoepp.aao.gov.au}
}
\altaffiltext{3}
{Max-Plank Institut fuer Astronomie, Koenigstuhl 17, \\ 
D-69117 Heidelberg, Germany \\
Electronic mail: {\tt ribata@mpia-hd.mpg.de}}
}

\begin{abstract}
Recently it has been suggested that the majority of dark matter in the
universe resides in  the form of Jupiter mass  black holes distributed
cosmologically.  This population makes itself apparent by microlensing
high  redshift  quasars and  introducing  pronounced variability  into
their observed light curves.   While several arguments dismissing this
hypothesis have  been presented,  a conclusive observational  test is,
alas, sadly  lacking.  In  this paper we  investigate the effect  of a
cosmologically  distributed  population   of  microlensing  masses  on
galaxies at low to intermediate redshift.  The magnification of bright
stars in  these galaxies leads to small,  but observable, fluctuations
in  their   surface  brightness.   The  variability   time  scale  for
Jupiter-mass lensing objects is of the  order of a few months and this
population  may be  detected through  a future  space-based monitoring
campaign of a field containing  $z \sim 0.5$ galaxies.  The monitoring
of galactic surface  brightness will provide an effective  test of the
nature of dark matter on cosmological scales.
\end{abstract}


\section{Introduction}\label{introduction}
The nature of dark matter is  one of the major outstanding problems in
current  astrophysics, with  the community  being divided  between two
broad  camps: those  who propose  particle dark  matter and  those who
support  massive  compact objects.   Both  these  ideas have  received
recent impetus due to  the measurement of a neutrino mass~\citep{fu99}
and the  detection of MACHOs  in the halo of  our Galaxy~\citep{al93},
although the  small mass  of the neutrino  and the uncertainty  in the
location of the  MACHO lenses suggests that the  nature of majority of
the dark matter still remains hidden from us.
 
One  recent suggestion  proposes  that the  variability  seen in  high
redshift quasars, rather than being intrinsic to the source, is due to
the  microlensing  action  of  Jupiter mass  black  holes  distributed
cosmologically    along    the    line-of-sight    to    the    quasar
\citep{ha93,ha96}.   To   account  for  the   characteristics  of  the
variability, a substantial population of black holes is required, with
a density  high enough  to account for  a significant fraction  of the
total mass  in the universe \citep{sc93}.  While  there is theoretical
support for such  a model, with models of  the quark-hadron transition
producing  black  holes  with   a  mass  of  $\sim10^{-3}{\rm  \msun}$
[e.g.~\citet{ha97}], observations have failed to reveal their presence
in     studies    of    the     equivalent    widths     of    distant
quasars~\citep{ca82,da91}, although this conclusion has been contested
\citep{ha96}.  Other  objections to this  microlensing hypothesis have
been   raised    \citep{ba95,al95},   with   corresponding   rebuttals
\citep{ha97}.   More recently, arguments  have become  quite vitriolic
with the suggestion that this idea of cosmologically distributed black
holes is being  ignored by the astronomical community  even though the
apparent evidence is overwhelming~\citep{ha99}.

The microlensing of cosmological supernovae provides a potential probe
of  any   population  of  black   holes  \citep{sc87,ra91,ra91a,me99},
although their rarity suggests that a long observational period may be
required until  an unambiguous  detection is made.   In this  paper we
provide a  complementary probe  of a cosmologically  distributed black
hole population;  instead of supernovae  as the source  population, we
focus here  on microlensing  on the much  more numerous  population of
normal   stars   in   galaxies    out   to   ${\rm   z\sim0.5}$.    In
Section~\ref{method} we outline the analytic and numerical approach to
this problem,  while in  Section~\ref{distributions} the results  of a
suite  of  simulations  are  presented.   Section~\ref{timescales}  is
concerned with the time scale of the microlensing variability, and the
observational   feasibility   of    this   study   is   discussed   in
Section~\ref{observational}.   Section~\ref{conclusions}  presents the
overall conclusions of this study.

\section{Method}\label{method}

\subsection{Background}\label{background}
Microlensing  surveys  of the  Galactic  halo  focus upon  identifying
brightness fluctuations of a  single, isolated source by an individual
star or  binary system  along the line-of-sight~\citep{pac86}.   It is
not  necessary, however,  to identify  single  stars in  a search  for
microlensing.   Proposing to  search for  MACHOs in  the halo  of M31,
Crotts (1992) noted that while  the light registered in a single pixel
is a combination of the  entire stellar population, the majority of it
arises in a relatively small number of extremely luminous giants. In a
stochastic manner  a massive object in  the halo of  M31 will traverse
the  line-of-sight  to  a  giant  and  induce  a  strong  microlensing
magnification, increasing the surface  brightness of the entire pixel.
A  monitoring program  of this  `pixel lensing'  has recently  come to
fruition  with the  identification of  several  potential microlensing
events~\citep{cr96,to96,al99}.

Given the relative lens geometry,  however, the optical depth for such
lensing is very  small (${\rm 10^{-6}\rightarrow10^{-5}}$), and events
are extremely  rare. Similarly, if all  dark matter is in  the form of
MACHO-like objects distributed uniformly through-out the universe, the
additional optical depth these would add in our view to M31 would, due
to  the  small  distance  and   hence  low  density  of  material,  be
negligible.  Focusing  instead out to  intermediate redshift, however,
the  optical  depth becomes  more  substantial  (this is  demonstrated
below) and a number of microlenses will lie along our line-of-sight to
a distant  galaxy. In such  a regime a  fraction of the  population of
stars  in a pixel  will be  subject to  strong magnification.   As the
intrinsic motions  of the microlensing objects  change their positions
with  respect to the  line-of-sight, the  `state' of  the microlensing
(i.e. how much each star  is magnified) will also change, resulting in
a fluctuation  in the  apparent luminosity of  a patch of  stars.  The
details of  these fluctuations  depends on several  factors, including
the optical depth to microlensing and the luminosity function of stars
in the source galaxy; these are explored in more detail below.

\subsection{Cosmological Considerations}\label{cosmology_c}
Standard cosmological models assume  that all material in the universe
is smoothly and homogeneously  distributed. In the following argument,
however, all  the matter  in the universe  is condensed in  to compact
objects.   To begin  with, it  is assumed  that these  objects  do not
influence our line-of-sight view to a distant source.  Dyer and Roeder
(1972;1973;1974) considered  the propagation of light  through such an
``empty beam''  universe and found that the  convergence, or focusing,
of the beam is removed due to the lack of material and that objects at
a particular redshift appear fainter than those observed in a universe
where the matter is smoothly distributed, even the universal curvature
is  the  same  in  both cases~\footnote{An  excellent  description  of
cosmological distances is given by Schneider, Ehlers \& Falco (1992)}.
For the purposes of this paper, we seek an intermediate position, with
matter  in the  form of  compact objects  distributed randomly  in the
universe, with  no constraint  on their location  with respect  to our
line-of-sight to a distant source.  With this, the microlensing masses
can influence  the beam  and, in the  mean, the appearance  of distant
objects must be  identical to that of a universe  where all the matter
is smoothly distributed \citep{we76}.   With respect to the empty-beam
universe model, the influence of  the lensing masses is to magnify the
distant sources by a factor
\begin{equation}
\langle\mu_{th}\rangle(z) = \left[\frac{r_1(z)}{r(z)}\right]^2~,
\label{meanamp}
\end{equation}
where $r(z)$ is  the normalized angular diameter distance  to a source
at redshift  $z$ in a universe  with a smooth  distribution of matter,
and $r_1(z)$  is the equivalent distance  in a universe  with the same
global  value of  $\Omega$, but  with all  the matter  in the  form of
compact  objects  that do  not  influence  the  line-of-sight view  to
distant objects.  These angular diameter distances are normalized such
that $D(z) = (c/H_o) r(z)$.

While the  smooth matter case is analytically  tractable, the solution
of   $r_1(z)$  typically   requires  the   numerical  solution   of  a
differential  equation  (Dyer \&  Roeder  1972;  1973;  1974). In  the
following, however,  we will assume  that $\Omega=1$ and that  all the
matter  is in  the  form  of substellar  compact  objects. With  these
assumptions~\citep{pac89};
\begin{equation}
\frac{r_1(z)}{r(z)} = 
\frac{3 + z^2 + 2\sqrt{ 1+z } + z(3+\sqrt{1+z})}
{5 ( 1 + z ) }~.
\label{distances}
\end{equation}
While this  tells us how bright  objects will appear in  the mean, the
distribution  of microlensing  masses along  any  random line-of-sight
will  be  different.  This  introduces  a  scatter  in the  brightness
distribution and, as the relative positions of the microlensing masses
and the  source will change  due to random  motions of the  lenses and
sources,  the magnification  along any  particular  line-of-sight will
vary with  time.  The first task  is to determine  the distribution of
magnifications for a population of microlensing masses.

\subsection{Microlensing}\label{microlensing}

\subsubsection{Microlensing Cross-Section}\label{crosssection}
To determine  the degree  of microlensing-induced fluctuations  on the
surface brightness  distribution of a distant galaxy,  it is necessary
to   first  derive   the  magnification   ${\rm   (\mu)}$  probability
distribution for  a source seen  through a population  of microlensing
masses. This task is simplified  by the fact that, irrespective of the
mass spectrum of the microlenses,  in the regime where $\mu$ is large,
$p(\mu)\propto  \mu^{-3}$~\footnote{Additional   caustic  features  do
modify  the  magnification probability  distribution  in this  region,
although  these   features  are  negligible  at   the  optical  depths
considered in this paper, a point we return to later.}.  In this case,
the distribution  is dominated  by bright pairs  of images and  can be
treated  analytically [e.g.~\citet{sc88}].   At  lower magnifications,
however,  the  form  of  the magnification  distribution  function  is
strongly dependent upon the  optical depth to microlensing, displaying
complex   secondary   features    due   to   higher   order   caustics
\citep{ra92,wa92,le95};  no  analytic  treatment fully  characterizing
these low  magnification features has been  presented [although recent
semi-analytical  attempts  have   proved  promising  e.g.   Kofman  et
al. (1997) \& Lee et al. (1997)].

For  the case  of a  single, isolated  microlens mass  the probability
distribution of magnifications can be derived analytically (Schneider,
Ehlers \& Falco 1992) and is given by
\begin{equation}
p_s(\mu)~d\mu \propto \frac{d\mu}{(\mu^2 - 1)^{\frac{3}{2}}}~.
\label{pmu}
\end{equation}
Note that this  probability is independent of the  mass of the lensing
object. It is  assumed that, in the regime relevant  to this work, the
microlensing optical depth  is small, and the masses  do not influence
one  another   to  produce  complex  caustic   patterns.   With  these
assumptions,  the  global  magnification probability  distribution  is
simply related to that of the isolated lensing mass.

\subsection{Normalization}\label{normalization}
While  Equation~\ref{pmu}  presents  the  form  of  the  magnification
probability distribution,  it still needs  to be normalized  before it
can employed in any analysis.   This question was addressed by Linder,
Wagoner \& Schneider  (1988) who determined that in  a universe with a
global density $\Omega$, of which a fraction $1-\alpha$ is in the form
of  compact  objects distributed  randomly  with  a constant  comoving
density, then
\begin{eqnarray}
& p(\mu,z) = & \frac{3}{2} \Omega ( 1 - \alpha )
\int^{z_s}_0 dz~\frac{(1+z)}{\sqrt{1+\Omega z}} 
\left[\frac{r_1(z)}{r_1(z_s)}\right]^2 \times \nonumber \\
& & \frac{r(z,z_s) r(z_s)}{r(z)} \int_0^{\infty} dm~\nu(m) p_s(\mu,m)~,
\label{linder}
\end{eqnarray}
where $\nu(m)$ is  the mass spectrum of the  lensing masses.  The last
integration  in this  expression reflects  the fact  that  the maximum
magnification a  source can undergo  is related to its  physical size,
the mass of the lensing objects and the lensing geometry~\citep{ch79}.
Here, $r(z_1,z_2)$ is the normalized angular diameter distance between
two redshifts.  For the gravitational lensing configurations presented
in  this paper,  assuming a  microlensing mass  of  $10^{-3}\msun$ and
considering   the   analysis   of   Chang  (1984),   $\mu_{max}   \sim
100\rightarrow200 / \sqrt{R}$,  where $R$ is the radius  of the source
star  in  Solar  radii.   The   most  luminous  B  and  A  stars  have
$R\sim10R_\odot$, for  which $\mu_{max}\sim30\rightarrow60$, while the
stars contributing the majority  of the populations luminosity possess
$R\sim1\rightarrow10R_\odot$.  This  upper limit of  the magnification
cuts off the probability distribution function, slightly enhancing the
distribution  below $\mu_{max}$  [e.g.  Figure  4 of  Lewis  and Irwin
(1995)].  As  significant magnification of stars can  occur, we chose,
therefore,  not  to restrict  the  probability distribution  function.
With this, Equation~\ref{pmu} is  independent of the microlensing mass
function, $\nu(m)$ can be replaced with an arbitrary $\delta$ function
and Equation~\ref{linder} can be  used to derive a normalization, $k$,
of the magnification probability distribution.

While    Equation~\ref{linder}   provides    a    normalization,   the
magnification    probability     distribution    as    described    by
Equation~\ref{pmu} diverges  as $\mu\rightarrow1$; this  is because in
determining  Equation~\ref{pmu} it  is assumed  that the  lensing mass
lies within an infinite plane,  with which any finite impact parameter
results in  $\mu > 1$.  An  infinite impact parameter  is required for
$\mu=1$.  As we are limited to  a finite amount of sky we require that
$p(\mu)\rightarrow0$ as  $\mu\rightarrow1$.  We choose,  therefore, to
modify Equation~\ref{pmu} with a  function to remove the divergence at
$\mu=1$.  The form employed is
\begin{equation}
p(\mu)   =    (   1    -   e^{\mu_w    (   1   -    \mu   )    }   )^2
\frac{k}{(\mu^2-1)^{\frac{3}{2}}}~,
\label{probability}
\end{equation}
where both the values $\mu_w$ and $k$ are functions of redshift.  This
Equation  is subject  to two  constraints, the  first being  the usual
normalization of the probability distribution function, and the second
being  flux  conservation  between   full  and  empty  beam  universes
(Equation~\ref{meanamp}):
\begin{eqnarray}
& & \int_1^{\infty} p(\mu)~d\mu = 1              \\
& & \int_1^{\infty} \mu~p(\mu)~d\mu = \langle \mu_{th} \rangle~.
\label{contraints}
\end{eqnarray}
However,  as Equation~\ref{linder} normalizes  the asymptotic  form of
the   probability   distribution,  the   single   free  parameter   in
Equation~\ref{probability}, namely  $\mu_w$, can be fixed  with one of
these  equations,   while  the  other   can  be  employed   to  ensure
consistency.

\begin{figure*}
\centerline{
\psfig{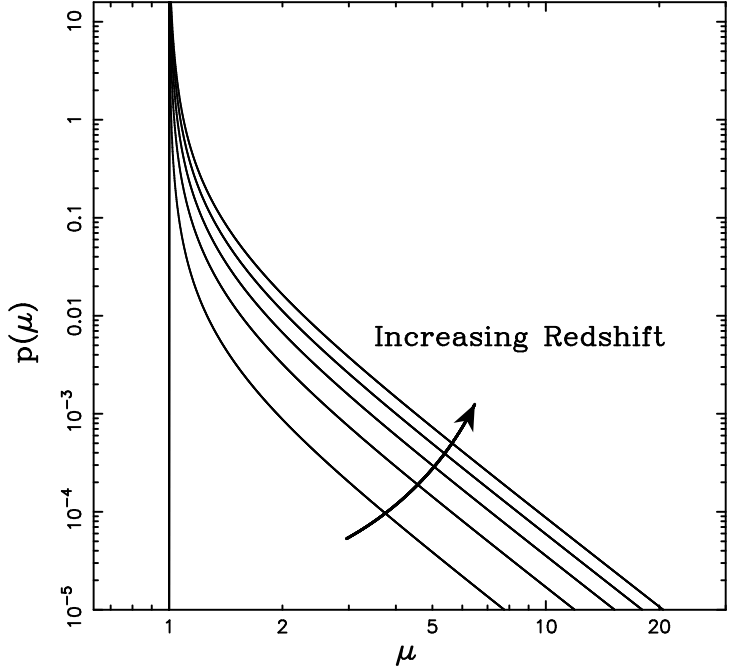}
}
\caption[]{  Microlensing magnification probability  distributions for
$z=$  0.1, 0.2,  0.3,  0.4 \&  0.5.   The corresponding  microlensing
optical  depths are $\kappa_*=$  0.002, 0.008,  0.017, 0.028  \& 0.040
respectively. }
\label{fig1}
\end{figure*}

\begin{figure*}
\centerline{
\psfig{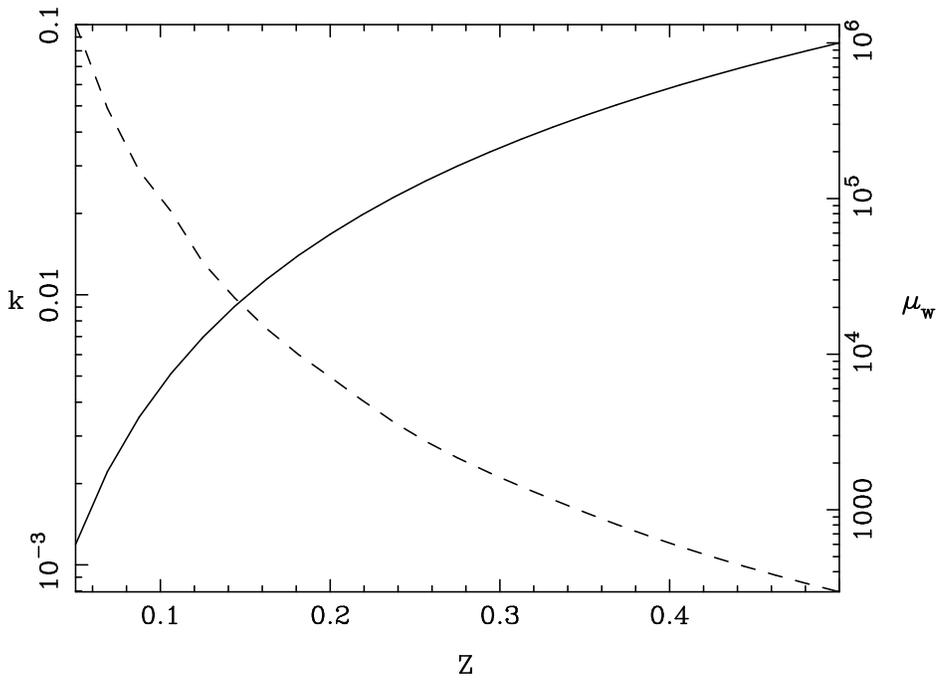}
}
\caption[]{ The parameters $k$  (solid line) and $\mu_w$ (dashed line)
for       the       magnification       probability       distribution
(Equation~\ref{probability})      in      the      redshift      range
$z=0.05\rightarrow0.5$. }
\label{fig1a}
\end{figure*}

\begin{figure*}
\centerline{ 
\psfig{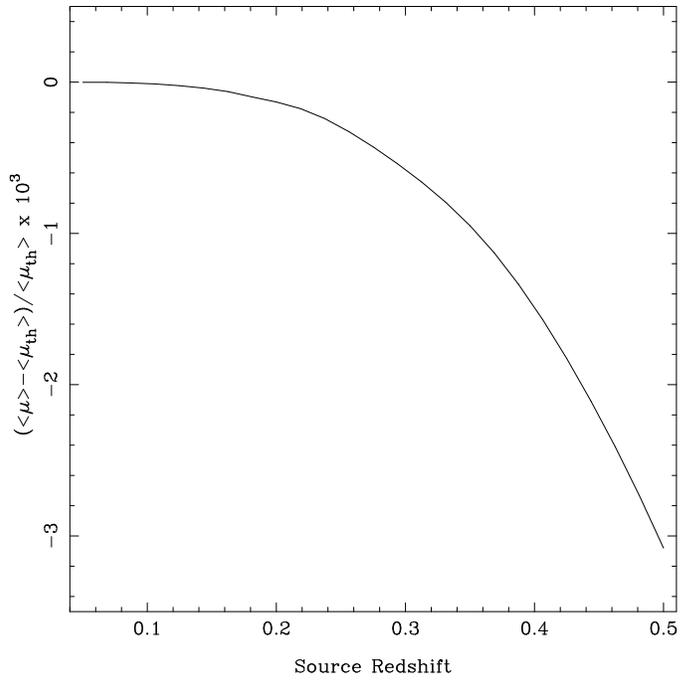}
}
\caption[]{ A  comparison between the  theoretical value of  the image
magnification, $\langle\mu_{th}\rangle$, compared to that derived from
the analytic approach presented in Section~\ref{method}.  }
\label{fig2}
\end{figure*}

The procedure employed in this paper is to use Equation~6 to determine
$\mu_w$.   A series  of  these probability  distributions, for  source
redshifts of  $z=$ 0.1, 0.2, 0.3,  0.4 \& 0.5, with  $\Omega=1$ all in
the form of compact objects, are presented in Figure~\ref{fig1}, while
Figure~\ref{fig1a} presents  $k$ and  $\mu_w$ over the  redshift range
$z=0.05\rightarrow0.5$.    All  of   the   magnification  probability
distributions  possess a  sharp  peak near  $\mu=1$,  coupled with  an
asymptotic  $p(\mu)\propto  \mu^{-3}$  tail  at  high  magnifications.
These normalized  distributions were  then used to  determine $\langle
\mu   \rangle$.   This   mean   magnification  is   compared  to   the
theoretically     expected    value     (Equations~\ref{meanamp}    \&
\ref{distances}) in Figure~\ref{fig2}; there is an excellent agreement
between the two approaches.  This is due to the low optical depth (and
hence simple) regime in which the analysis is undertaken, as at higher
stellar densities,  such as those responsible for  the microlensing of
multiply  imaged  quasars,  individual  stars no  longer  behave  like
isolated  lenses and  their  combined effects  can  lead to  extremely
complex magnification patterns on a high redshift source~\citep{wa90}.

Considering the  theoretically expected magnification it  is simple to
determine that the  optical depth to microlensing out  to the redshift
of  interest.   This ranges  from  $\kappa_*\sim0.002$  at $z=0.1$  to
$\kappa_*\sim0.04$ at  $z=0.5$.  These  numbers are several  orders of
magnitude  greater  than  the  $\kappa_*\sim10^{-6}\rightarrow10^{-5}$
expected  of  gravitational  microlensing  in the  Galactic  halo  and
towards  M31 [c.f.~\citep{pac86}],  and hence,  in  this extragalactic
scenario, microlensing  effects should be more  apparent.  While these
arguments  demonstrate  the  efficacy  of microlensing,  deriving  the
detailed  behavior  requires   knowledge  of  the  underlying  stellar
luminosity function  as the number  density of potential  sources will
dictate the level of observed  fluctuations in the stellar sample as a
whole, a topic which we turn to in the next section.

How does Equation~\ref{probability} compare to previous studies of the
magnification probability  distribution of microlenses  distributed in
three dimensions?  The numerical  studies undertaken by Rauch (1991;a)
focus on substantially higher  optical depths than those considered in
this paper, although one case, with $\kappa_*=0.01$ [Figure 2 in Rauch
(1991)],    can   be   compared    to   distribution    presented   in
Figure~\ref{fig1}   for  $z=0.2$  $(\kappa_*=0.008)$.    An  excellent
correspondence  between the  two distributions  can be  seen  over the
entire  magnification range.   Kofman et  al.  (1997)  and Lee  et al.
(1997) analytically tackled the  form of the magnification probability
distribution  for  both  2-D  and 3-D  distributions  of  microlensing
masses. Also  considering the numerical simulations  of Rauch (1991;a)
and Rauch  et al. (1992), these  papers investigate the  nature of the
higher  order   caustic  features  identified   in  the  magnification
probability distribution  at large magnifications,  features which are
not accounted for  in Equation~\ref{probability}.  While substantially
influencing  the form of  the 2-D  distributions, these  features were
found to be  very small for the corresponding  3-D distributions.  The
analysis  of Kofman  et al.   (1997) and  Lee et  al.  (1997)  was not
considered  at the  optical depths  presented in  this paper,  but for
$\kappa_*=0.1$  the  caustic  features  induce  a  deviation  of  only
$\sim15\%$  at  $\mu\sim30$,  where  the distribution  has  fallen  to
$10^{-6}$ that of its peak  value. At the optical depths considered in
this paper, such deviations will be even more inconsequential, and can
be ignored. Lee et al. (1997) comment on difficulty in determining the
form   of   the   magnification   probability  distribution   at   low
magnifications, especially  in the case where  the microlensing masses
are distributed in  three dimensions. Comparing Figure 6  of Kofman et
al.  (1997) with Figure  13 of Lee et al. (1997), it  can be seen that
at small  magnifications ($log_{10}(\mu) < 0.1$)  that the probability
distributions  for the 2-D  and 3-D  case are  very similar.   In this
regime, Kofman  et al. (1997) provide a  semi-analytic formalism which
approximates  the  form  of   the  distribution.   Comparing  this  to
Equation~\ref{probability}   for  the   $z=0.5,   \kappa_*=0.04$  case
presented  in  this  paper,  a  maximum  difference  of  $\sim7\%$  at
$log_{10}(\mu)\sim  0.005$.  Hence  Equation~\ref{probability}  has an
excellent correspondence with previous results.

\subsection{Stellar Luminosity Function}\label{lumfun}
With  current instrumentation,  the  angular diameter  of galaxies  at
redshifts  of $z\sim0.5$  ensures that  they will  extend over  only a
small number  of pixels, and the  light recorded in each  pixel is the
sum of the flux from a large  number of stars.  Each star in the pixel
will be  microlensed by some degree,  and examining Figure~\ref{fig1},
most stars will suffer a magnification $\mu\sim1$, although a fraction
of the population will be magnified by a much more substantial factor.
On average, however, the overall flux  in a pixel will be magnified by
$\langle\mu_{th}\rangle$, although at any  instant the brightness of a
pixel will  depend on  the relative positions  of the  microlenses, as
this dictates how much each star in the pixel is magnified. These will
change due  to the intrinsic  velocity dispersion of  the microlensing
bodies, resulting in fluctuations in  the observed flux within a pixel
about this mean value. To be a useful probe of the intervening matter,
however, these fluctuations must exceed some observable threshold.  To
properly determine this, the luminosity function of the source stellar
population needs to be considered.

If  the  luminosity  function  is  comprised of  only  relatively  low
luminosity stars then a large  number is required to account for total
flux detected in  a pixel. When such a  population is microlensed many
stars will be in a  highly magnified state.  As the microlensing state
changes, and the magnification of each star also changes, the absolute
number  of stars  in the  highly magnified  state  remains essentially
constant. Even  if a single star  is extremely magnified,  a very rare
occurrence,  then  the  contribution  this  star  makes  to  the  flux
registered in a pixel is small compared to the overwhelming background
of   the  numerous   stars  which   are  only   moderately  magnified.
Conversely, if the luminosity  function consists of only very luminous
stars  then  only a  small  number are  required  to  account for  the
luminosity in a  pixel.  It is very unlikely  that any individual star
is strongly magnified.  When this  rare event does occur, however, the
contribution from the magnified star can dominate the flux received in
a  pixel,  resulting in  a  substantial  fluctuation  in the  observed
surface brightness.  The real  situation, however, is a combination of
the above and  we expect that a stellar population  sampled in a pixel
consists of a large proportion of faint stars which are insensitive to
the effects  of microlensing,  resulting in a  background flux,  and a
relatively  small number  of very  luminous stars  which  dominate the
luminosity.

The  luminosity  function model  we  choose  is  the stellar  LF  from
Jahrei\ss\  and Wielen  (1997),  derived from  Hipparchos data  within
$20\pc$ of  the Sun.  The characteristics of  this luminosity function
are  presented in  Figure~4; representing  a population  with  a total
luminosity of  $10^6\lsun$, the solid  line shows the number  of stars
possessing a luminosity greater  than some value, while the dot-dashed
line is their  contribution to the total luminosity  of the pixel.  It
is  obvious  from  this  picture  that while  stars  with  $L>10\lsun$
represent less than 1\% of  the population by number, they account for
70\% of the total luminosity.

\section{Magnification Distributions}\label{distributions}
Given  a population  of microlensing  masses, the  magnification  of a
single  star  can  be  chosen  by  selecting  from  the  magnification
probability distribution  given by Equation~\ref{probability}.   For a
sample of  stars, magnifications can  be drawn from  this distribution
multiple times to determine the total magnification of the population.
While this  is straight forward, it  becomes computationally expensive
when  the population  consists  of  a very  large  number stars.   The
procedure  employed in this  study is  to determine  the magnification
statistics for various numbers  of source stars, combining the results
to  reproduce the  effect of  microlensing the  entire  source stellar
population.   Analytically this  can be  calculated by  convolving the
magnification   probability    distribution   for   a    single   star
(Equation~\ref{probability})  with   itself  $n_{samp}$  times,  where
$n_{samp}$ is the number of stars being microlensed. Again, while this
is  simple for  a small  population of  source stars,  the calculation
becomes numerically  unwieldy for  a large number  of stars.   For the
purpose of this  paper, these statistics were generated  a Monte Carlo
approach,       drawing       $n_{samp}$       observations       from
Equation~\ref{probability}  and combining them  to determine  the mean
magnification of  the sample.   Repeating this procedure  $10^6$ times
for each $n_{samp}$ it is  then possible to determine the distribution
of  mean magnifications.   The results  of this  sampling  for several
source  redshifts  between  $z=0.05$  and  $z=0.5$  are  presented  in
Figure~\ref{fig4}, with  sample sizes ranging from 1  to $10^4$ source
stars. At all  redshifts the same trend is  seen with the distribution
of   the   means   evolving    from   the   single   realization   for
(Equation~\ref{probability})  to a  more Gaussian  form as  the sample
size is increased; this is a consequence of the central limit theorem.
These   distributions  can   then   be  combined   to  determine   the
distributions of means for any sample size.

Armed with this information it  is now possible to tackle the question
of the  overall brightness fluctuations  expected for a  population of
stars drawn from a luminosity function, and hence the expected surface
brightness  fluctuation   within  a   pixel.   For  this,   the  above
magnification distributions are  convolved with the stellar luminosity
function   described  in   Section~\ref{lumfun},   assuming  a   total
luminosity, and  hence number  of stars in  the population.   Again, a
Monte Carlo  approach is taken  and the luminosity function  is binned
logarithmically. While the bins at lower luminosity contain many stars
they contribute a negligible fraction  of the total luminosity and are
uniformly magnified by $\left<\mu\right>$. At the high luminosity end,
where  the  luminosity  contributions   of  the  bins  become  a  more
appreciable   fraction   of   the  total   luminosity,   magnification
probability distribution  for each bin is calculated  by combining the
appropriate   population   magnification   probability   distributions
presented in Figure~\ref{fig4}.  The distribution for each bin is then
combined, weighted by the luminosity fraction of the bin, to determine
the magnification  probability distribution for  the entire population
of  stars.   Two  constraints,  that  the  mean  value  of  the  final
distribution must  be $\left<\mu\right>$ and that  the integrated area
under  the  resultant  probability  distribution must  be  unity,  are
checked  to  ensure   consistency.   Figure~\ref{fig4a}  presents  the
results  of this  procedure for  a range  of source  redshifts between
$z=0.05$  and  $z=0.5$,  displaying  the  percentage  variability  for
stellar  populations of $10^5\lsun$  (dark grey),  $10^6\lsun$ (medium
grey)    and    $10^7\lsun$    (light    grey)    total    luminosity.
Table~\ref{table1}   presents  the   integral   properties  of   these
distributions with each column representing the range within which 10,
30,  50, 70  \&  90\% of  the  magnifications of  the population  lie,
relative  to  the  average  magnification.  A  number  of  interesting
features  are immediately  apparent;  at low  redshift the  relatively
small microlensing optical  depth induces negligible fluctuations into
the observed total brightness of any of the populations.  This changes
as  the  redshift,  and  hence  the  microlensing  optical  depth,  is
increased,  until  by  redshift  $z=0.5$ microlensing  will  introduce
fluctuations of order $\sim1\%$ into  a sample with a total luminosity
of  $10^5\lsun$.  Increasing  the  luminosity of  the stellar  sample,
however, reduces the  width of the distribution, as  expected from the
central limit  theorem, such  that a population  of $10^7$  at $z=0.5$
will   suffer  fluctuations   of   $\simlt0.5\%$.  We   turn  to   the
observational     aspects    of     such    small     variations    in
Section~\ref{observational}.

\section{Event Timescales}\label{timescales}
While the magnification probability distribution is independent of the
form  of  the  underlying   mass  spectrum  of  lensing  objects,  the
time-scale of  any fluctuation depends  implicitly on the mass  of the
microlenses.   At   the  low  optical  depths   considered  here,  any
variability  will   be  very  simple   in  form,  consisting   of  the
superposition of  isolated peaks rather than the  complex light curves
expected for high  optical depth microlensing [e.g.~\citet{le93}].  As
with the  microlensing of  Magellanic Cloud sources,  we can  define a
variability time-scale to be:
\begin{eqnarray}
\tau & = & \frac{(1+z_l)}{v} \left[ \frac{4GM}{c^2}
\frac{D_{ls}D_{ol}}{D_{os}}\right]^{\frac{1}{2}} \nonumber \\
& \approx & 23.4 \frac{(1+z_l)}{v_{1000}}
\left(\frac{M}{M_\odot}\right)^{\frac{1}{2}}
\left[\frac{r(z_l,z_s) r(z_l)}{r(z_s)}
\right]^{\frac{1}{2}}~h^{-\frac{1}{2}}~yrs~,
\label{einstein}
\end{eqnarray}
where  $M$  is  the  microlensing  mass, $v_{1000}$  is  the  relative
transverse velocity  in units of  $1000\kms$, and $r(z_1,z_2)$  is the
angular diameter distance.

For   source   galaxies  out   to   $z\sim0.5$,   and  considering   a
$10^{-3}\msun$  microlensing  mass, this  time-scale  is presented  in
Figure~\ref{fig5}.   The  time-scale of  the  microlensing events  are
substantially  less than  a year  for all  lensing  configurations and
hence can  be observed in  a single observing season.   When examining
Figure~\ref{fig5}, it  is important to remember  that the distribution
of microlenses are  volume-weighted (c.f.  Equation~\ref{linder}), and
hence  are   more  likely   to  be  located   nearer  to   the  source
redshift. Similarly the typical time-scale for the microlensing events
seen in the  light curve of a population of stars  will also be skewed
to values  less than the maximum seen  in Figure~\ref{fig5}.  Detailed
light curves are beyond the scope  of this current paper and they will
be the subject of a further study of this phenomenon.

\section{Observational Considerations}\label{observational}

\subsection{Variability Detection}\label{variability}
While the  previous theoretical  analysis has presented  the framework
for  determining   the  effects   of  a  cosmological   population  of
microlensing masses, it is important to consider whether the amplitude
of  the   expected  variations  are  observable.    Here  we  consider
monitoring many  distant galaxies simultaneously with  a CCD detector,
with resolution  elements that each  cover a population  of luminosity
$\sim  10^7\lsun$.   Ignoring K-corrections,  this  corresponds to  an
apparent  magnitude of  $m=29.0$  at $z=0.5$  (assuming $q_0=0.5$  and
$h=0.75$).   It  is  immediately  clear that  such  observations  will
normally  be  sky-limited,  except   for  the  Next  Generation  Space
Telescope (NGST) for  which the sky background is  expected to be very
low, and the spatial resolution high enough to have very small pixels.

To  detect 1\%  variations  at the  $3\sigma$  level, 0.33\%  relative
photometry  is needed, that  is, ${\rm  S/N =  333}$.  Assuming  an 8m
diameter mirror,  and a  diffraction limited system,  a count  rate of
$\sim 1.8  S_{eff}$~photons/sec is expected from a  $m=29.0$ object in
the J-band with NGST, where  $S_{eff}$ is the total system efficiency.
For exposure times  longer than a few minutes,  such observations will
be    photon    noise     dominated,    since    a    background    of
$<10^{-1}$~photons/sec/resolution~element
\footnote{see  http://www.ngst.stsci.edu/sky/sky.html} is  expected at
$1.1\mu m$. So  the required $S/N$ is achieved  in $17/S_{eff}$~hrs of
exposure.  For a population of  stars at a redshift of $z\sim0.5$ with
a total  luminosity of $10^7\lsun$,  the probability of  the necessary
1\% variation is 4\%, so  several thousand such populations need to be
monitored to detect a sample of microlensing events.  (Although deeper
observations,  focusing upon  the  fainter regions  of galaxies,  will
reveal similar scale fluctuations  in $\sim23\%$ and $\sim56\%$ of the
pixels for populations of, respectively, $10^6\lsun$ and $10^5\lsun$).
Of course,  in a  typical field, many  resolution elements on  the CCD
camera will cover $z \sim 0.5$  galaxies, and so the monitoring of the
populations  may be  achieved simultaneously.   If NGST  is resolution
limited, each resolution element  will cover $0.001 \Box{\scmd}$.  For
each resolution element to cover a $10^{7}\lsun$ population at $z \sim
0.5$, requires the observation of a region of surface brightness ${\rm
21.5   mag/\Box{\scmd}}$.   In   regions  fainter   than   ${\rm  21.5
mag/\Box{\scmd}}$,  boxes  larger than  a  resolution  element can  be
summed up, to give a  total luminosity $10^7\lsun$; the variability of
these boxes can also be studied  with the penalty of a small amount of
extra sky noise.

The HDF  field shows that  the area in  a typical $2'\times  2'$ field
covered by  surface brightness ${\rm >  21.5 mag/\Box{\scmd}}$ regions
of $z=0.5$  galaxies exceeds several  tens of $\Box{\scmd}$,  that is,
tens of thousands of NGST resolution elements.  The above cosmological
microlensing-induced variability will be observed over a background of
intrinsic variability  events, due  to supernovae, variable  stars and
galaxy self-lensing, as well as due to the inevitable Poisson noise in
the observed  counts.  Unlike  current studies of  `surface brightness
fluctuations' method  for distance  determinations out to  galaxies at
$\sim100Mpc$~\citep{Tonry1988,Thomsen1997,Lauer1998} which essentially
requires single  epoch observations, to  identify microlensing induced
variability   a   monitoring  program   is   required,  allowing   the
identification   of  supernovae   by  their   light-curves   and  peak
luminosities. Supernovae are  also rare, as are variable  stars with a
luminosity  sufficient enough  to  influence the  total luminosity  of
populations  $L\sim10^7L_\odot$, if  a  survey was  to  focus on  more
luminous source  pixels.  As with supernovae, these  would also reveal
themselves via their characteristic light curves.  If the depth of the
survey is  increased such  that $L\sim10^5L_\odot$ pixels  are probed,
the ubiquity of their variations  would rule against variations in the
source population.   In the  case of self-lensing  of galaxy  stars by
MACHOs  in  the  halo  of   the  same  galaxy  provides  a  negligible
contribution as the optical depth  is many orders of magnitude smaller
than that  due to the  cosmological microlenses considered  here.  The
main  source of  detected background  events is  likely to  be  due to
simple  Poisson  photon noise.   However,  in  the situation  outlined
above, $3\sigma$ variations due  to Poisson noise are approximately 40
times less common than the $3\sigma$ microlensing variations.  Data at
two epochs would allow an  initial feasibility study.  If the required
variability rate is indeed observed, follow-up observations at further
epochs should be taken to  monitor the light curves of the variability
events, to distinguish microlensing  events from unlensed supernova or
nova events, and to reject variations due to Poisson noise.

\subsection{Extreme Events}\label{extreme}
\begin{figure*}
\centerline{
\psfig{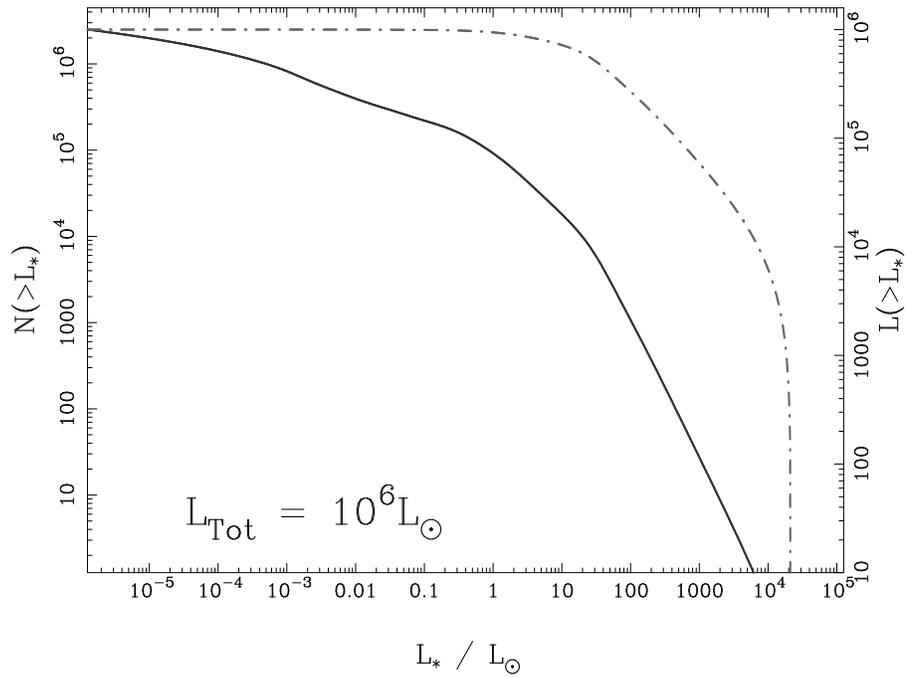}
}
\caption[]{The properties of the assumed luminosity function, assuming
a population with  a total luminosity of ${\rm  10^6\lsun}$. The solid
line  represents  the  number  of  stars  in  the  population  with  a
luminosity greater  that ${\rm L_*}$,  while the dot-dash line  is the
summed luminosity  of this  subset of stars.   While stars  with ${\rm
L>10\lsun}$ represent only a small fraction of the over all population
by  number  they  are  responsible  for  the  majority  of  the  total
luminosity.  }
\label{fig3}
\end{figure*}

\begin{figure*}
\centerline{
\psfig{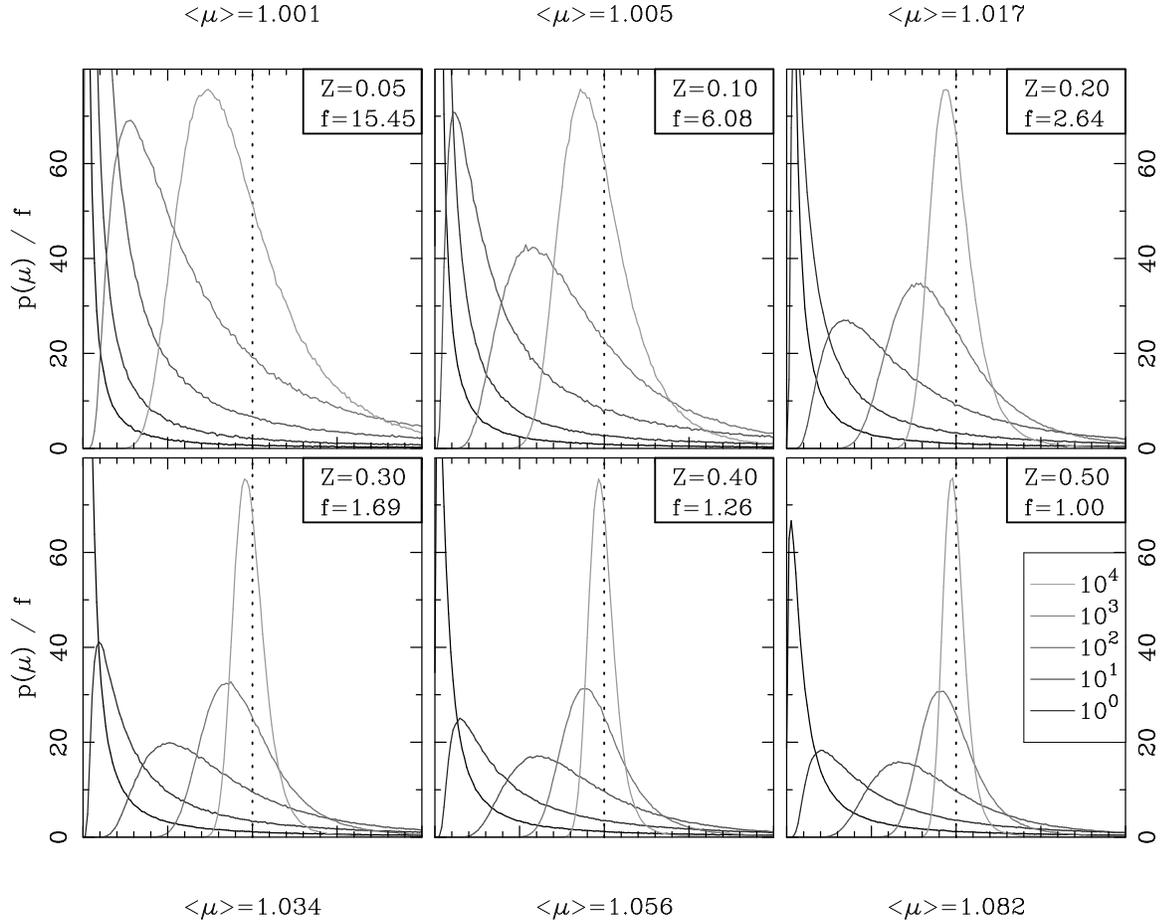}
}
\caption[]{The distribution  of the mean magnification of  a sample of
stars for redshifts between  $z=0.05$ and $z=0.5$. The dotted vertical
line  represents   the  mean  value  of   the  relevant  magnification
distribution, while the lower bound on each plot is $\mu=1$. Each line
consists of  a stellar  sample of  1, 10, 100,  1000 and  10000 stars,
depending on its shading. The scale factor, f, is introduced such that
all  the curves  can be  placed  on the  same y-scale.   As these  are
probability  distributions,  the   integrated  area  under  the  curve
(accounting for the f-scaling) is unity.
}
\label{fig4}
\end{figure*}

\begin{figure*}
\centerline{
\psfig{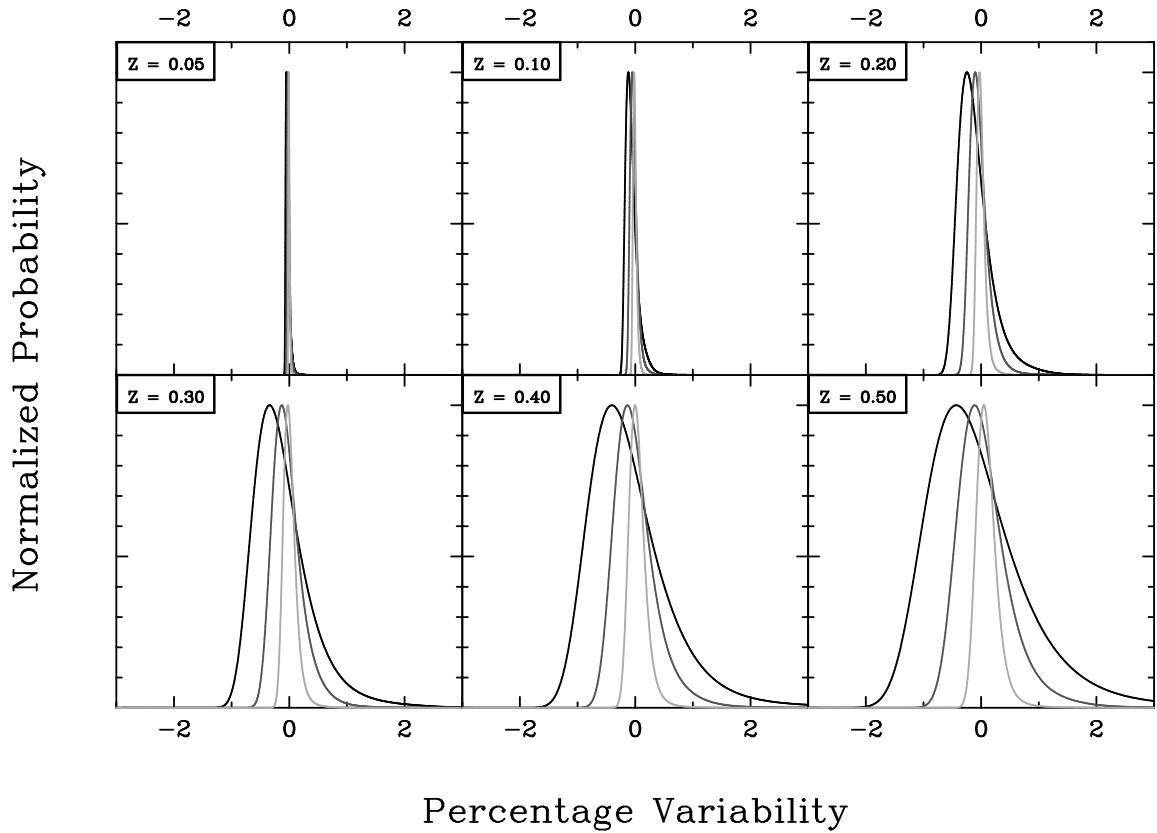}
}
\caption[]{The  distribution in  total luminosity  of  stellar samples
from $10^5\lsun$  (broadest - dark  grey) to $10^7\lsun$  (narrowest -
light grey), defined  by the luminosity function in  the text. At very
low  redshifts,  the  low  microlensing  optical  depth  induces  only
negligible variability,  while by a redshift  of $z=0.5$, fluctuations
of order  a percent  are induced. The  integrated properties  of these
distributions are given in Table~\ref{table1}.}
\label{fig4a}
\end{figure*}

\begin{figure*}
\centerline{
\psfig{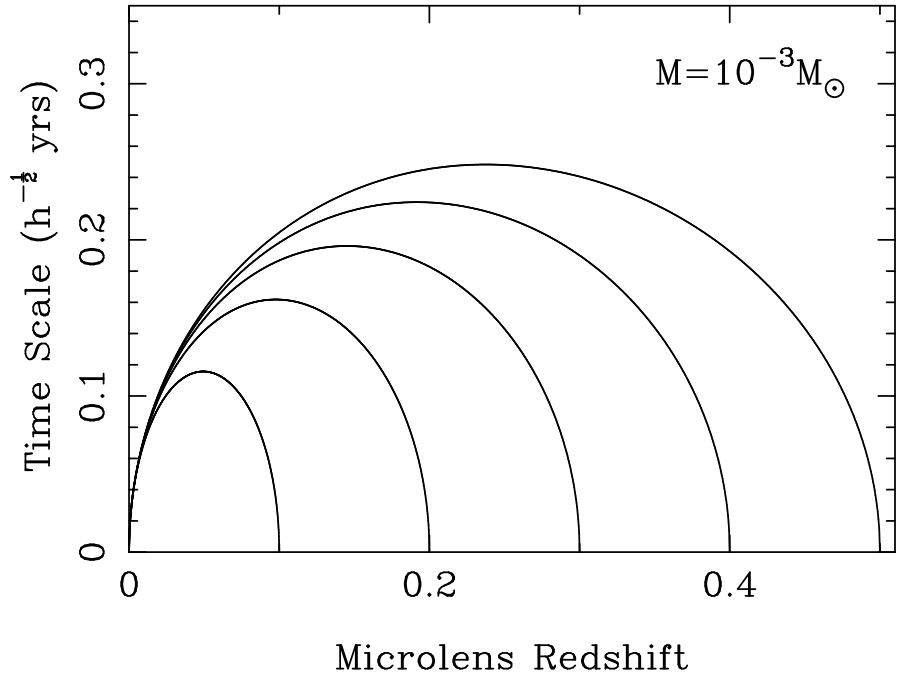}
}
\caption[]{The   variability  time   scale   for  a   $10^{-3}M_\odot$
microlensing mass  with a relative transverse  velocity of 1000$\kms$,
for sources at $z=$ 0.1, 0.2, 0.3, 0.4 \& 0.5~.}
\label{fig5}
\end{figure*}

\begin{figure*}
\centerline{
\psfig{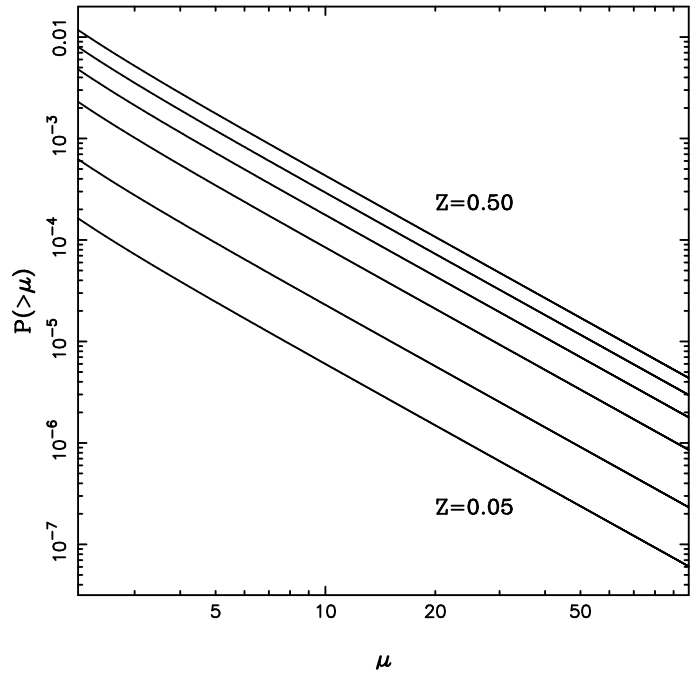}
}
\caption[]{The cumulative  magnification probabilities for  sources at
$z=$ 0.05, 0.1, 0.2, 0.3, 0.4 \& 0.5~.}
\label{fig6}
\end{figure*}

The previous Section has demonstrated that the typical fluctuations in
the brightness  of a  pixel is detectable  and lends itself  to future
space-based observatories.  But are  there any aspects of microlensing
by  a  cosmologically  distributed  population that  would  make  them
apparent  with current technology?   As noted  earlier in  this paper,
while the vast majority of stars will be magnified by a value close to
the theoretically expected  mean (Equation~\ref{meanamp}), very rarely
a star will undergo an  extreme magnification.  The probability that a
particular  star  is  magnified  by  an  extreme  value  is  found  by
integrating     the     magnification     probability     distribution
(Equation~\ref{probability}),   which  is  presented   graphically  in
Figure~\ref{fig6}.  For instance, in  a $10^5\lsun$ population, the LF
of Figure~\ref{fig3}  gives $\sim 0.1$ stars with  $L \sim 10^4\lsun$;
if such a star were to be  magnified by a factor of 20, it would alter
the surface brightness of the population by $200$\%.  At a redshift of
$z=0.05$, this  is an uncommon phenomenon,  with probability $P(>\mu)=
10^{-6}$, however,  the apparent magnitude  of the population  is much
brighter:  $m=27.5$, and  the photometric  accuracy needed  to resolve
these variations  is ${\rm S/N \sim  3}$.  While these  are within the
range  of HST's  capabilities, and  many tens  of fields  of  depth of
$m=27.5$ with ${\rm S/N \simgt  3}$ have been reimaged, it is unlikely
that a  sufficient area has  been covered to  detect a sample  of such
rare events. So, while the  tools are available, more substantial deep
survey areas are required in a search for extreme events with HST.

\section{Other Cosmological Models}\label{cosmology}
\begin{figure*}
\centerline{
\psfig{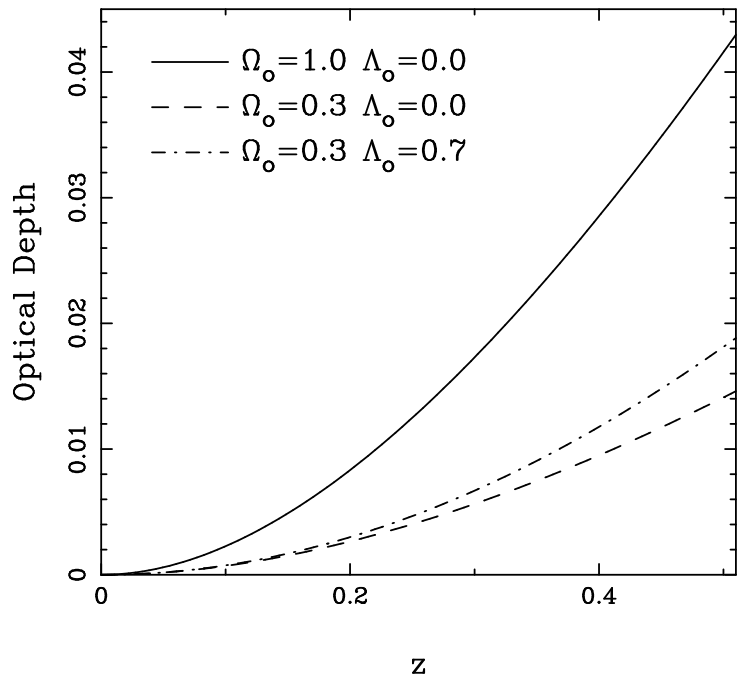}
}
\caption[]{The optical depth out to $z=0.5$ in several cosmologies. At
the  low  redshifts considered,  the  introduction  of a  cosmological
constant has only a small influence on the optical depths.}
\label{fig7}
\end{figure*}

While   throughout  this   paper  a   standard  flat   cosmology  with
$\Omega_o=1$  has  been  employed,  recent studies  of  high  redshift
supernovae  suggest   the  presence  of   a  substantial  cosmological
constant~\citep{ga98}.   To  investigate   the  influence  of  such  a
dominant cosmological  constant on the result presented  in this paper
the formalism of light  propagation in generalized Friedmann universes
was employed~\citep{Linder1988}. The full  and empty beam distances in
three  universes   were  calculated  in   three  cosmological  models,
$(\Omega_o=1,\Lambda_o=0)$,    $(\Omega_o=0.3,    \Lambda_o=0)$    and
$(\Omega_0=0.3,\Lambda_o=0.7)$,  used to  calculate the  mean relative
magnification    between    the   full    and    empty   beam    cases
(Equation~\ref{meanamp}) and the microlensing optical depth. These are
presented in Figure~\ref{fig7}.

Out  of  these  three  models,   the  case  presented  in  this  paper
consistently displays  higher optical  depth over the  entire redshift
range; this is simply due to the fact that in this model more material
lies along the line of sight  to a distant source.  In comparison, the
optical  depth in  the open  case is  several times  smaller, reaching
$\sigma\sim0.15$ at $z=0.5$.  The  addition of a dominant cosmological
constant into  such a universe does  not greatly change  the values of
the optical  depths. This is because, while  the cosmological constant
changes the global  curvature of the Universe, it  has no influence on
the Ricci focusing of the beam  and locally the Universe appears as it
is solely  matter-dominated (it is  for this reason  that cosmological
supernovae  need to  be identified  at redshifts  greater than  0.5 in
an attempt to determine the geometry of the Universe).

\section{Conclusions}\label{conclusions}
Recent  studies have  claimed that  the fluctuations  observed  in the
light  curves of  high redshift  quasars are  due to  the microlensing
effect  of a  population  of cosmologically  distributed Jupiter  mass
black holes. While arguments have been presented to refute this claim,
no observational  test of this hypothesis has  been readily available.
In this paper we have investigated the effect this putative population
would have on  the surface brightness distribution of  galaxies out to
intermediate  redshift.   The   results  of  this  study  conclusively
demonstrate  that  if   a  significant  population  of  cosmologically
distributed compact  objects are present, with enough  mass to account
for  the  universal  dark  matter  budget, then  they  will  induce  a
`twinkling'  into  the observed  surface  brightness distributions  of
galaxies  at low  to  intermediate redshifts.   The  magnitude of  the
observed fluctuations  are a  function of the  redshift of  the source
galaxy,  ranging from  negligible  values in  the  local Universe,  to
$\sim1$\% at $z=0.5$, with a time-scale of variability on the order of
weeks to months.

While the  optical depth to  microlensing at such low  to intermediate
redshifts is still  quite small and the induced  fluctuations are at a
relatively  low  level, the  sky  density  of  galaxies out  to  these
redshifts vastly  exceeds that of  supernova and quasars,  the current
focus  for  the  search  for cosmologically  distributed  microlensing
masses.  With sufficiently deep imaging with source pixel luminosities
of greater than $10^5L_\odot$, microlensing induced surface brightness
variability can  be detected over galaxies  at intermediate redshifts,
and the ubiquitous  nature of this variability means  it will dominate
over any contaminating effects, such as supernovae, variable stars and
self-lensing.  The scale of  the induced variability make observations
conducive over a  single observing season, and with  the advent of the
Next  Generation  Space  Telescope,  such observations  will  soon  be
possible.  Hence,  a monitoring program to search  for fluctuations of
the  surface brightness of  the plethora  of galaxies  at $z\simlt0.5$
offers a effective test of the existence of cosmologically distributed
compact matter.

\section{Acknowledgements}
We  are grateful  for  discussions  with Mike  Hudson,  Tom Quinn  and
Stephen Gwyn. The anonymous referee is thanked for useful comments.

\newpage

\newpage

\newpage

\newpage

\newpage

\newpage

\newpage

\newpage

\newpage

\newpage

\begin{deluxetable}{cccccccc}
\tablecolumns{8}  
\tablewidth{0pc}  
\tablecaption{}
\tablehead{  
\colhead{}    &  \colhead{} &   \colhead{}   &  
\colhead{10\%} & \colhead{30\%} & \colhead{50\%} & \colhead{70\%} & \colhead{90\%}}
\startdata  
         & $10^5{\rm L_\odot}$ && $\pm$0.010 & $\pm$0.027 & $\pm$0.040 & $\pm$0.052 & $\pm$0.065  \\
 z=0.05  & $10^6{\rm L_\odot}$ && $\pm$0.004 & $\pm$0.013 & $\pm$0.021 & $\pm$0.029 & $\pm$0.041  \\
         & $10^7{\rm L_\odot}$ && $\pm$0.002 & $\pm$0.005 & $\pm$0.010 & $\pm$0.014 & $\pm$0.021  \\ \hline
         & $10^5{\rm L_\odot}$ && $\pm$0.021 & $\pm$0.063 & $\pm$0.101 & $\pm$0.138 & $\pm$0.186  \\
 z=0.10  & $10^6{\rm L_\odot}$ && $\pm$0.010 & $\pm$0.031 & $\pm$0.051 & $\pm$0.075 & $\pm$0.108  \\
         & $10^7{\rm L_\odot}$ && $\pm$0.003 & $\pm$0.013 & $\pm$0.022 & $\pm$0.033 & $\pm$0.052  \\ \hline
         & $10^5{\rm L_\odot}$ && $\pm$0.047 & $\pm$0.139 & $\pm$0.235 & $\pm$0.335 & $\pm$0.480  \\
 z=0.20  & $10^6{\rm L_\odot}$ && $\pm$0.022 & $\pm$0.070 & $\pm$0.122 & $\pm$0.177 & $\pm$0.266  \\
         & $10^7{\rm L_\odot}$ && $\pm$0.009 & $\pm$0.027 & $\pm$0.046 & $\pm$0.071 & $\pm$0.116  \\ \hline
         & $10^5{\rm L_\odot}$ && $\pm$0.072 & $\pm$0.215 & $\pm$0.365 & $\pm$0.530 & $\pm$0.786  \\
 z=0.30  & $10^6{\rm L_\odot}$ && $\pm$0.036 & $\pm$0.104 & $\pm$0.187 & $\pm$0.277 & $\pm$0.427  \\ 
         & $10^7{\rm L_\odot}$ && $\pm$0.010 & $\pm$0.040 & $\pm$0.070 & $\pm$0.108 & $\pm$0.175  \\ \hline
         & $10^5{\rm L_\odot}$ && $\pm$0.092 & $\pm$0.285 & $\pm$0.490 & $\pm$0.720 & $\pm$1.094  \\
 z=0.40  & $10^6{\rm L_\odot}$ && $\pm$0.047 & $\pm$0.143 & $\pm$0.252 & $\pm$0.373 & $\pm$0.578  \\ 
         & $10^7{\rm L_\odot}$ && $\pm$0.016 & $\pm$0.052 & $\pm$0.088 & $\pm$0.137 & $\pm$0.233  \\ \hline
         & $10^5{\rm L_\odot}$ && $\pm$0.112 & $\pm$0.352 & $\pm$0.608 & $\pm$0.899 & $\pm$1.377  \\
 z=0.50  & $10^6{\rm L_\odot}$ && $\pm$0.049 & $\pm$0.168 & $\pm$0.305 & $\pm$0.459 & $\pm$0.715  \\
         & $10^7{\rm L_\odot}$ && $\pm$0.005 & $\pm$0.057 & $\pm$0.108 & $\pm$0.159 & $\pm$0.279  
\enddata  
\tablecomments{\label{table1}The integrated properties of the probability distributions given
in Figure~\ref{fig4a}. The columns presents the range over which 10, 30, 50, 70 \& 90\% of the
magnifications are found, relative to the mean value (i.e for a $10^6{\rm L_\odot}$ population
at z=0.40, 50\% of all magnifications lie within $\pm0.252\%$ of the mean value).}
\end{deluxetable}

\end{document}